\documentclass[10pt, letterpaper, conference]{IEEEtran}
\IEEEoverridecommandlockouts
\usepackage{cite}
\usepackage{amsmath,amssymb,amsfonts}
\usepackage{graphicx}
\usepackage{textcomp}
\def\BibTeX{{\rm B\kern-.05em{\sc i\kern-.025em b}\kern-.08em
    T\kern-.1667em\lower.7ex\hbox{E}\kern-.125emX}}

\usepackage{balance}
\usepackage{pifont}
\usepackage{framed}
\usepackage{soul}
\usepackage{multirow}
\usepackage{array}
\usepackage{algorithm2e}
\RestyleAlgo{ruled} 
\usepackage{fontawesome}
\usepackage{graphicx}
\usepackage{booktabs}
\usepackage{mathtools}
\usepackage{acronym}
\usepackage[shortlabels]{enumitem}
\usepackage[export]{adjustbox}
\SetKw{Continue}{continue}
\newcolumntype{L}[1]{>{\raggedright\let\newline\\\arraybackslash\hspace{0pt}}m{#1}}
\newcolumntype{C}[1]{>{\centering\let\newline\\\arraybackslash\hspace{0pt}}m{#1}}
\newcolumntype{R}[1]{>{\raggedleft\let\newline\\\arraybackslash\hspace{0pt}}m{#1}}
\usepackage{url}

\usepackage[caption=false,font=footnotesize,labelformat=simple,listofformat=subsimple]{subfig} 

\usepackage[bookmarks=false,hidelinks]{hyperref} 

\usepackage[svgnames]{xcolor}

\usepackage[frozencache]{minted} 
\makeatletter
\let\@float@c@listing\@caption
\makeatother

\usepackage{csquotes}
\MakeOuterQuote{"}

\acrodef{CPS}{Cyber-Physical System}
\acrodef{IoT}{Internet of Things}
\acrodef{HDL}{Hardware Description Language}
\acrodef{CAD}{Computer-Aided Design}
\acrodef{EDA}{Electronic Design Automation}
\acrodef{HPC}{High-Performance Computing}
\acrodef{DL}{deep learning}
\acrodef{ML}{machine learning}
\acrodef{NLP}{natural language processing}
\acrodef{IC}{Integrated Circuit}
\acrodef{CWE}[CWE]{Common Weakness Enumeration}
\acrodef{CVE}[CVE]{Common Vulnerabilities and Exposures}
\acrodef{LLM}[LLM]{large language model}
\acrodef{NMT}[NMT]{neural machine translation}
\acrodef{HLS}[HLS]{High-Level Synthesis}
\acrodef{SVA}[SVA]{SystemVerilog Assertion}
\acrodef{AST}[AST]{Abstract Syntax Tree}
\acrodef{IP}[IP]{Intellectual Property blocks}
\acrodef{HT}[HT]{Hardware Trojan}
\acrodef{HLL}[HLL]{High-Level Language}
\acrodef{RTL}[RTL]{Register-Transfer level}
\acrodef{PPA}[PPA]{Power, Performance, and Area}
\acrodef{SoC}[SoC]{System-on-Chip}
\acrodef{isa}[ISA]{Instruction Set Architecture}
\acrodef{sdl}[SDL]{Secure Development Lifecycle}
\acrodef{roi}[ROI]{Return on Investment}
\acrodef{apr}[APR]{automatic program repair}

\newlist{inlinelist}{enumerate*}{1}
\setlist[inlinelist]{label=(\roman*), itemjoin={{, }}, itemjoin*={{, and }}}
    
\begin{document}
\bstctlcite{IEEEexample:BSTcontrol}

\title{
An Investigation of Hardware Security Bug Characteristics in Open-Source Projects
}

\author{
\IEEEauthorblockN{Joey Ah-kiow}
\IEEEauthorblockA{\textit{Center for Cybersecurity} \\
\textit{NYU Tandon School of Engineering} \\
}
\and
\IEEEauthorblockN{Benjamin Tan}
\IEEEauthorblockA{\textit{Department of ESE} \\
\textit{University of Calgary}\\
}
}

\maketitle

\begin{abstract}
Hardware security is an important concern of system security as vulnerabilities can arise from design errors introduced throughout the development lifecycle. 
Recent works have proposed techniques to detect hardware security bugs, such as static analysis, fuzzing, and symbolic execution.
However, the fundamental properties of hardware security bugs remain relatively unexplored. 
To gain a better understanding of hardware security bugs, we perform a deep dive into the popular OpenTitan project, including its bug reports and bug fixes. 
We manually classify the bugs as relevant to functionality or security and analyze characteristics, such as the impact and location of security bugs, and the size of their bug fixes.
We also investigate relationships between security impact and bug management during development.
Finally, we propose an abstract syntax tree-based analysis to identify the syntactic characteristics of bug fixes. 
Our results show that 53\% of the bugs in OpenTitan have potential security implications and that 55\% of all bug fixes modify only one file.
Our findings underscore the importance of security-aware development practices and tools and motivate the development of techniques that leverage the highly localized nature of hardware bugs. 

\end{abstract}


\section{Introduction}
\label{sec:intro}
The security of hardware is paramount to computer system security. 
Modern systems are increasingly complex and reliant on hardware-based security features to ensure the confidentiality, integrity, and availability of information. 
Errors in these features can have serious implications, and verifying the security of designs as part of a ``security development lifecycle'' (SDL) is crucial~\cite{intel_sdl_2022}. 

In the SDL, the early \textit{architecture stage} involves the development of security requirements and objectives using threat modeling. 
High-level security requirements are refined and translated into design specifications at the \textit{design stage}.
Errors in these two stages could result in under-specified behavior and introduce security bugs during implementation.
Hardware designers can introduce security bugs during the \textit{implementation stage} due to implementation errors from misinterpreted specifications, incorrect assumptions, typos, etc.
Current processes are largely manual, rely on human expertise, and suffer from many limitations when faced with the complexity and hardware-software interactions of modern \ac{SoC} designs~\cite{Dessouky_hardfails_2019}.

Recent work proposes myriad techniques for improving the security verification of hardware, including fuzzing~\cite{Chen_HyPFuzz_2023}, property generation~\cite{Deutschbein_isadora_2021}, and symbolic execution~\cite{Zhang_coppelia_2018}. 
Notably, static code analysis~\cite{Ahmad_cweat_2022} focuses on security scanning of designs written in hardware description languages (HDLs), like Verilog, in the earliest stages of implementation, offering a highly desirable \textit{shift left} approach.
However, it relies on assumptions of bug characteristics; for instance, that they are are highly localized and can be detected statically on a file-by-file basis, without additional context from other design elements (i.e., pre-elaboration). 

\Ac{apr} also relies on assumptions about bug characteristics; CirFix~\cite{Ahmad_cirfix_2022} proposed a technique using design repair templates based on language constructs commonly found in hardware (functional) bug fixes identified in prior work~\cite{Sudakrishnan_verilog_bugfix_patterns_2008}. 
It is unclear if the common constructs are similar in hardware security bugs. 
Thus, we seek a better understanding of the characteristics of hardware security bugs in behavioral HDL, as these may be useful to guide new bug detection and repair approaches.

In this paper, we draw from open-source hardware projects to investigate hardware security bug characteristics. 
While prior work examined general Verilog bug fixes in a limited study~\cite{Sudakrishnan_verilog_bugfix_patterns_2008} and the code redundancy assumption~\cite{Xu_He_Zhang_Yang_Wu_Mao_2023}, we are the first to provide a deep dive into hardware security bugs specifically. 
We examine several open-source hardware projects to explore existing hardware security-related bug/issue reporting and perform a deeper dive into a security-centric system-on-chip design. 
We analyze various dimensions of security bugs and their associated bug fixes, including their impact, location, and footprint.

We also implement an AST-based approach to identify finer-grained characteristics of bug fixes that can inform future work in static analysis scanners.
Our contributions include the following:

\begin{itemize}
    \item Analyses of 235 bug reports from a complex, open-source SoC project
    \item Characterization of hardware security bugs in HDL based on features such as impact and location that have not been studied before to our knowledge
    \item Insights into the corresponding bug fixes, including AST-level characteristics
    \item Open-source data and analysis tools to facilitate further research
\end{itemize}

\autoref{sec:prior} further motivates our study and briefly discusses prior related work. 
\autoref{sec:experiments} details our study and research questions, including different ways to characterize bugs and fixes. 
\autoref{sec:discussion} provides additional discussion and takeways, and \autoref{sec:conclusions} concludes. 

\section{Motivation and Related Prior Work\label{sec:prior}}
Bugs are commonly considered implementation errors that may result in product behavior that deviates from its desired intent.
These deviations from intended behavior are referred to as \textit{failures}~\cite{Avizienis_dependable_secure_computing_2004}.
Failures occur due to one or more \textit{faults}.
For our purpose, bugs are a subset of faults -- All bugs are faults, but not all faults are bugs.
This deviates from the definitions used by Tan et al.~\cite{Tan_bug_characteristic_oss_2014}, who used the terms "bug", "defect", and "fault" interchangeably.
This is a necessary distinction, as hardware faults can occur for various reasons, both intentionally (e.g., fault injection attack) and unintentionally (e.g., manufacturing faults), outside of its implementation. 
This distinction also clarifies the relationship between bugs and \textit{hardware Trojans} (HTs); HTs are intentionally introduced malicious modifications, unlike bugs which are unintentional.

Bugs are not always caused during the implementation step. 
The development lifecycle is essentially a pipeline of iterative translations, and any error in that pipeline can introduce bugs.
Design intent is gradually refined throughout development and molded into different shapes -- natural language specifications in the initial stages, \ac{RTL} designs during implementation, and tests during validation -- all of which must agree on the correct behavior. 
Errors in the design stage, such as under-specified behavior, have as much potential to introduce bugs as errors during implementation do.

\subsection{Security Bugs, Weaknesses, and Vulnerabilities}
It is important to differentiate between security bugs, CWEs, and vulnerabilities. 
A CWE is a ``weakness'' found in software, firmware, hardware, or service component that may contribute to the introduction of vulnerabilities~\cite{cwe_intro_2023}. 
A vulnerability is one or more weaknesses that are exploitable by an adversary. 
Vulnerabilities are the exploitable subset of security bugs, and security bugs are the "design-related" subset of CWEs.
A security bug is a bug that violates the intended confidentiality, integrity, or availability of a design.
Confidentiality, integrity, and availability are commonly thought of as the tenets of information security and other security objectives, such as authentication, are derivatives of these three.

Security bugs can be more challenging to address than functional bugs~\cite{Dessouky_hardfails_2019}, partly because security analysis needs speculation of possible malicious intent, and are thus context sensitive (i.e., threat model).
For example, let us consider a function $y = f(x)$. 
A functional requirement is that $y$ has the expected value for all values of $x$. 
A potential security requirement is that the time it takes to compute $y$ is constant for all values of $x$.
The functional requirement is easier to specify and thus verify simply because the value of $y$ is the purpose of the computation, whereas the security requirement is counter-intuitive and contradicts performance goals.
This contention between design goals (i.e., performance vs. security) is an additional layer of complexity whose resolution is highly reliant on expertise and context (threat model). 
The functional requirement could also be a security one, depending on the context in which the computation is used; if $y$ is an asset (i.e., something that is critical to proper behavior\footnote{\url{https://csrc.nist.gov/glossary/term/asset}}), then the integrity of $y$ is a security requirement.

The relationship between bugs and vulnerabilities is \textit{many-to-many};
a vulnerability may be the result of one or more bugs, and a bug may be attributed to one or more vulnerabilities. 
The relationship between bugs and CWEs is similar.
The security implication becomes more concrete as we transition from CWEs to bugs and from bugs to vulnerabilities.
Security bugs are more concrete than CWEs because they contain specific implementation details and context. They contain characteristics of one or more CWE. 
Vulnerabilities are more concrete than bugs because there is a specific attack vector used that leverages bug(s). 
On one hand, techniques like fuzzing~\cite{Kande_thehuzz_2022, Chen_HyPFuzz_2023} and symbolic execution~\cite{Zhang_coppelia_2018} identify vulnerabilities -- bugs and their associated exploits; these high-effort/high-reward approaches provide the most concrete results. 
On the other hand, static code analysis~\cite{Ahmad_cweat_2022} attempts to find as many \textit{potential} problems as possible, as early and cheaply as possible. However, this approach relies on manual expertise to filter and validate potential errors.

As we discuss in the next section, to the best of our knowledge, there are few studies that examine the source code characteristics of hardware security bugs and fixes in detail. 
Exploring these characteristics, especially in the context of a mature open-source hardware project, can provide insights for ongoing and future research. 
The 
\subsection{Related Prior Work}
Sudakrishnan et al.~\cite{Sudakrishnan_verilog_bugfix_patterns_2008} studied the bug fix patterns of four Verilog designs available on OpenCores. 
They determined patterns in bug fix syntactical characteristics and grouped them into seven major categories: \begin{inlinelist}
    \item if-related
    \item module-declaration
    \item module-instantiation
    \item assignment
    \item switch
    \item always
    \item class field.
\end{inlinelist}
They found that assignment-related patterns are the most common (29-55\%), followed by if-related patterns (18-25\%), module declaration/instantiation (4.7-25\%), class-field-related patterns (4.7-16\%), always-related (2.8-21\%), and case-related patterns (1.8-4.6\%).
This study provided insights into the language constructs most associated with general hardware bugs. However,
the complexity of the designs investigated and characteristics studied are limited. For example, they do not study how many lines or files are changed for in the bug fix. 

More recently, Hardfails~\cite{Dessouky_hardfails_2019} explored \textit{software-exploitable hardware security bugs}, 
identifying four limitations of current automated verification techniques: 
\begin{inlinelist}
    \item Cross-modular effects
    \item Timing-flow gap
    \item Cache-state gap
    \item Hardware-software interactions.
\end{inlinelist}
Their work featured a mix of ``native'' bugs and those ``inserted'' as part of the Hack@DAC contest. 
In contrast, we attempt a more quantitative study to observe the fundamental characteristics of hardware security bugs from a source code perspective.

Xu et al.~\cite{Xu_He_Zhang_Yang_Wu_Mao_2023} validated the redundancy assumption for hardware. 
They studied 1405 bug fixes across 12 open-source Verilog projects and found that 17.71\% and 11.77\% of donor code for \ac{apr} can be found within the global (project) scope and local (file) scope, respectively. 
They also found that processes, identifier definitions, and assign statements are the most common statements in the identified code clone pairs. 
While HDL bug analysis is less mature, there are security-specific analyses for software. 
Zaman et al.~\cite{Zaman_Adams_Hassan_2011} completed a case study of the Firefox project. 
Tan et al.~\cite{Tan_bug_characteristic_oss_2014} studied three large open-source project and obtained 1087 security bugs from the NVD database.
In this work, we propose approaches to gain similar insights for HDL projects. 

\section{Experimental Work and Results \label{sec:experiments}}
\subsection{Research Questions}
To gain a better understanding of hardware security bugs in open-source hardware projects, we analyze both the bug reports and their associated bug fixes. 
We guide our characterization through the following research questions (RQs):
\begin{enumerate}
    \item[RQ1.] {Bug Characteristics: 
        \begin{enumerate}
            \item {How many bugs are security related?}
            \item {What is the impact of hardware security bugs?} 
            \item {How are security bugs managed in development?}
        \end{enumerate}
    }
    \item[RQ2.] {Security Bug Fix Characteristics:
        \begin{enumerate}
            \item {What is the size/locality of bug fixes for security bugs?}
            \item {What language constructs are commonly associated with bug fixes?}
        \end{enumerate}
    }
\end{enumerate}

\subsection{Project Evaluation}
Prior works relied on the CVE and NVD databases to obtain a reliable dataset of software security bugs~\cite{Tan_bug_characteristic_oss_2014, Wei_comprehensive_sec_bug_study_2021}. 
For our work, there is not enough data available for open-source hardware projects in such databases, so we established our bug dataset by identifying projects with a suitable corpus of bugs. 
The development of high quality open-source hardware such as the one make 

We initially considered the 10 popular open-source RISC-V-based designs shown in \autoref{tab:osh} and extracted their bug reports from the start of their lifetime to 2023-05-25.
All 10 projects use GitHub issues for bug reporting. 
We identified relevant issues by filtering for closed issues and used issue labels to filter for RTL bugs. 
We focused on closed issues because they contain more contextual information (e.g., developers’ discussion, bug fix) required to identify the bug type.
Four projects did not have the necessary labels to complete this step and were discarded.
Of the remaining projects, OpenTitan presents itself as the most informative design, given the following:
(i) it makes up the majority of reported bugs (58\%) across all projects, 
(ii) it offers the most comprehensive collateral (e.g., documentation, firmware, etc.)
(iii) it is the most complex design, integrating multiple high-quality peripherals, and 
(iv) it is a Root-of-Trust \ac{SoC} with an extensive suite of security countermeasures. 
(v) the sample size of bugs in other projects is too small to obtain meaningful project-level insights

Finally, we manually inspected the 235 RTL bug reports from OpenTitan and removed 65 because we could not find their bug fixes or they were not obviously design bugs (i.e., DV bugs, build bugs, false reports). 
The final dataset contained 170 bugs.

\begin{table}[!t]
\caption{Open-source RISC-V Designs and Issues/Bugs Reported in Their GitHub Issues}
\label{tab:osh}
\centering
\scriptsize
\resizebox{0.8\columnwidth}{!}{%
\begin{tabular}{cccc}
\toprule
\textbf{Project} & \textbf{\# of Issues} & \textbf{\# of Bugs} & \textbf{\# of RTL Bugs (\% of bugs)} \\ \midrule
\textbf{OpenTitan} & 4148 & 516 & 235 (45.5) \\ 
\textbf{OpenPiton} & 22 & N/A & N/A \\ 
\textbf{Ibex} & 603 & 128 & 49 (32.3\%) \\ 
\textbf{CVA6} & 469 & 45 & 20 (44.4\%) \\ 
\textbf{BlackParrot} & 82 & N/A & N/A \\ 
\textbf{NeoRV32} & 21 & 8 & N/A \\ 
\textbf{CV32E41P} & 3 & N/A & N/A \\ 
\textbf{CV32E40S} & 80 & 38 & 23 (84.2\%) \\ 
\textbf{CV32E40X} & 153 & 45 & 28 (62.2\%) \\ 
\textbf{CV32E40P} & 348 & 102 & 52 (51\%) \\ \bottomrule
\end{tabular}%
}
\vspace{-1em}
\end{table}

\subsection{Insights into Bug Characteristics}
\noindent \textbf{RQ1.a) How many bugs are security related?}
To classify bugs into security and functional bugs, we manually analyzed all 170 bug reports and the associated bug fixes to identify an impact on confidentiality, integrity, and availability in the context of OpenTitan's threat model, to the best of our understanding. 
In cases where the bug report was not enough to properly understand the type of the bug, we also considered the corresponding bug fix.
If a clear security impact could not be determined, the bugs were labeled as functional bugs by default.

We found that 50.0\%, 57.7\%, 54.8\%, and 49.3\% of bugs per year between 2018 to 2022 were security bugs.
Overall, \textbf{52.9\% of all bugs studied were security bugs}. 
This reveals security bugs make up a significant portion of bugs in OpenTitan and there have been no changes in the proportion of security bugs across bugs reported during its lifetime. 
To put this in perspective, 35\% of bugs in the Linux kernel in 2011
were security bugs~\cite{Tan_bug_characteristic_oss_2014}.
This result reaffirms the importance of hardware security to overall system security.\\

\noindent \textbf{RQ1.b): What is the perceived impact of security bugs?} 
For each security bug found, we identified its impact on confidentiality, integrity, and availability;
in cases where a bug has multiple impacts, it is counted in all applicable impacts.
We determined that \textbf{34 (37.8\%), 47 (52.2\%), and 43 (47.8\%) have an impact on confidentiality, integrity, and availability, respectively}. 
This result aligns with expectations as confidentiality and availability concerns typically manifest in very specific contexts, whereas integrity issues can occur due to any functional error involving assets.

\begin{figure}[t]
    \centering
    \includegraphics[width=0.8\columnwidth]{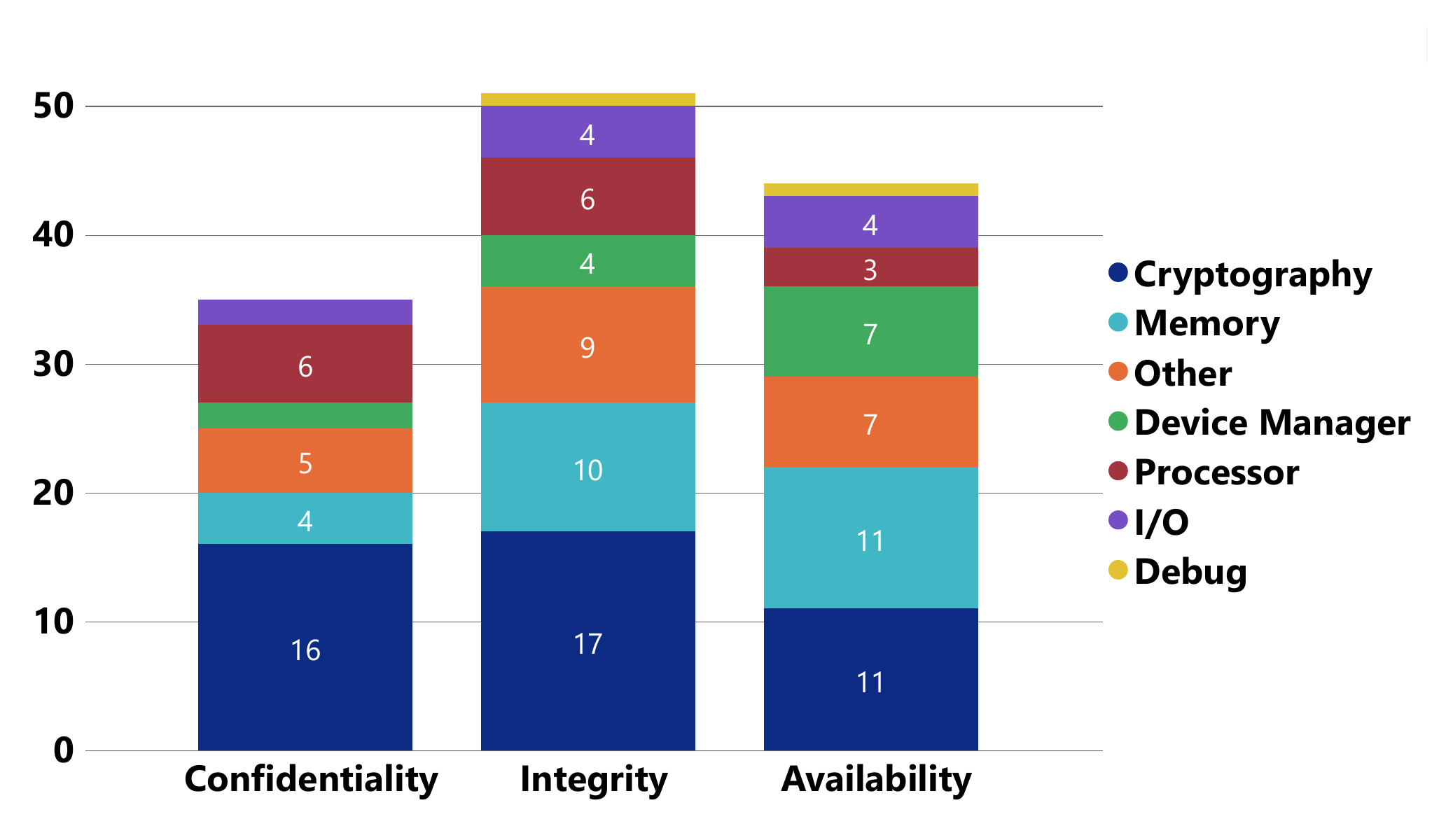}
    \vspace{-1em}
    \caption{\# Bugs categorized by Impact and IP Category}
    \vspace{-1.5em}
    \label{fig:cia_location}
\end{figure}

We also investigate the relationship between the impact and location of the bug. 
We defined the location as the IP block that was modified in the associated bug fix. 
For our analysis, we grouped each IP into categories that represent their high-level functionality (shown in \autoref{tab:ip_category}). 
\autoref{fig:cia_location} illustrates the number of bugs categorized by their impacts and locations. 
As before, bugs with multiple impacts and/or locations are counted in all applicable impacts and locations. 
The \textbf{IP category with the most security bugs was cryptography}.
We also observed a higher correlation between confidentiality and cryptography than other impacts/locations; 45.7\% of confidentiality bugs were in cryptography IPs, compared to 33.3\% and 25\% for integrity and availability, respectively.
These results are sensible because cryptography IPs are security-critical by nature and any error within are more likely to cause security concerns, and cryptography is largely used to ensure the confidentiality of information, compared to integrity and availability which can be impacted due to many factors.\\

\begin{table}[t]
\centering
\caption{IP blocks in OpenTitan and their our categorization}
\label{tab:ip_category}
\resizebox{0.7\columnwidth}{!}{%
\begin{tabular}{@{}cl@{}}
\toprule
\textbf{IP Category} & \multicolumn{1}{c}{\textbf{IP}} \\ \midrule
Cryptography & aes, keymgr, hmac, kmac, csrng, edn \\ 
Memory & flash\_ctrl, otp\_ctrl, rom\_ctrl \\ 
I/O & spi\_device, spi\_host, pinmux \\ 
Device Manager & rstmgr, pwrmgr, clkmgr, sysrst\_ctrl \\ 
Processor & otbn, ibex \\ 
Debug & rv\_dm \\ 
Other & tlul, xbar, prim, aon\_timer, alert\_handler \\ \bottomrule
\end{tabular}%
}
\vspace{-1.5em}
\end{table}

\noindent \textbf{RQ1.c): How are security bugs managed?}
We also investigate how security bugs are managed during development and look at two dimensions: \begin{inlinelist}
    \item {the number of messages (developers' discussion) involved in a bug report and its bug fix}
    \item {the number of days it takes to resolve a bug}
\end{inlinelist}.

These give us insight into how much contextual information is shared during the bug-fixing process and the prioritization of which type of bugs to fix, respectively. 

\begin{figure}[t]
    \centering
    \includegraphics[width=0.9\columnwidth]{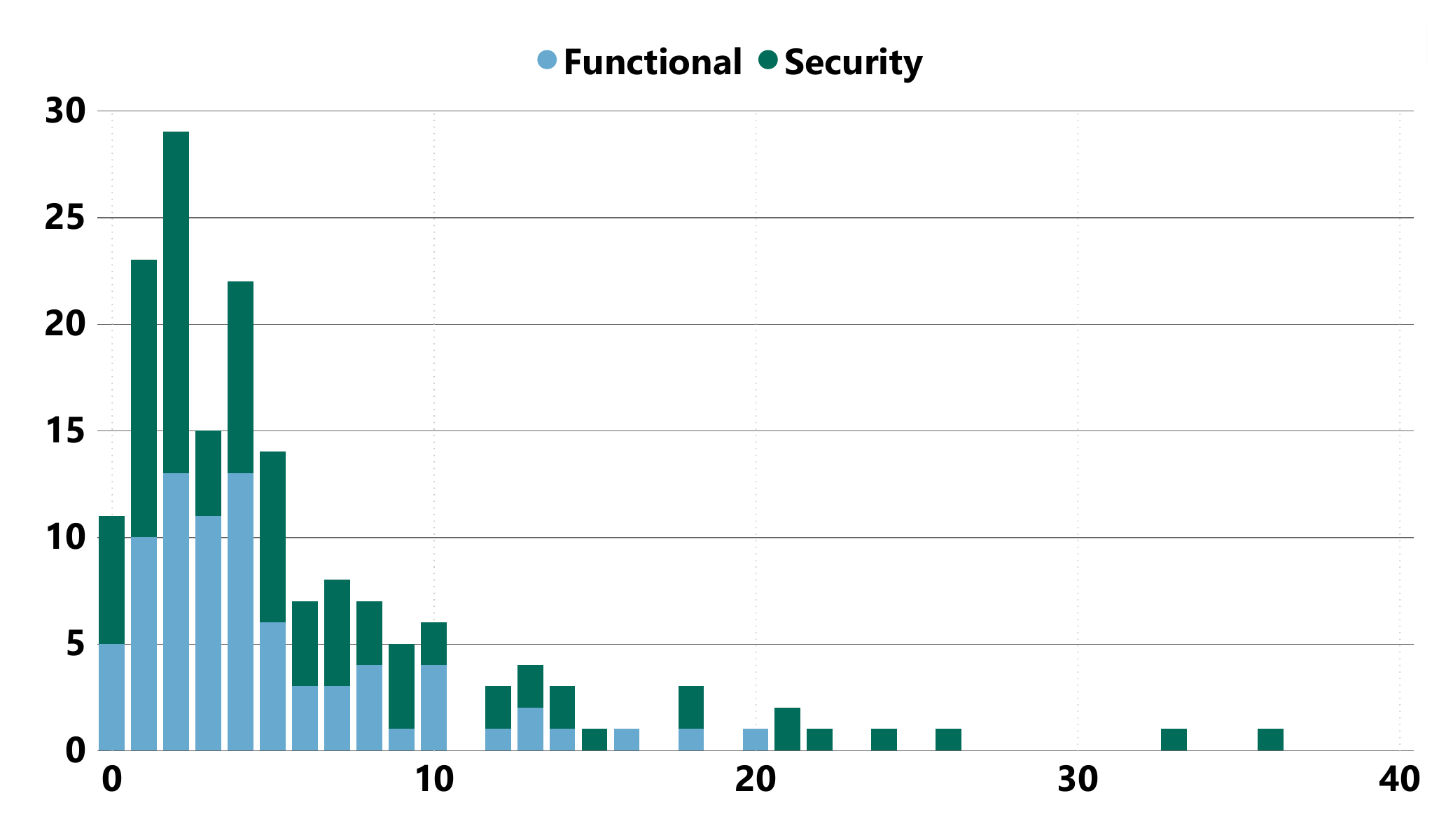}
    \vspace{-1.5em}
    \caption{\# Bugs vs. \# messages in the bug reports. (Distribution of the number of messages in reports and their associated fixes.)}
    \vspace{-1.5em}
    \label{fig:num_comments}
\end{figure}

\begin{figure}[t!]
    \centering
    \includegraphics[width=0.9\columnwidth]{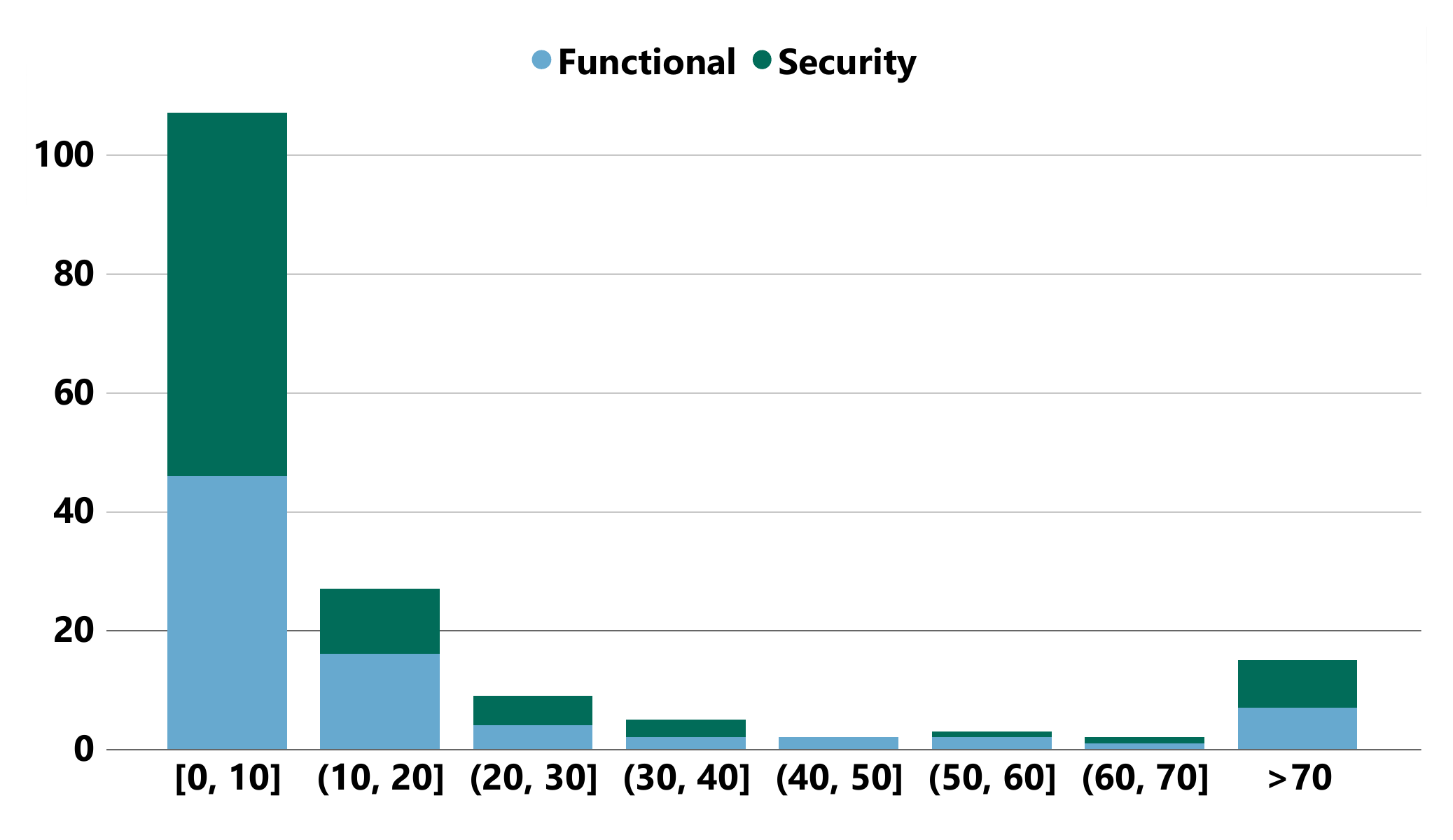}
    \vspace{-1.5em}
    \caption{\# Bugs fixed vs. \# Days required (Distribution of the number of days required to close bug reports). Days are grouped in increments of 10.}
    \vspace{-1.5em}
    \label{fig:time_to_fix}
\end{figure}

The ``\# of messages'' comprises the message count in the bug report (issue) and bug fix (pull request), excluding the initial message made when creating the issue/pull request. 
The distribution is shown in \autoref{fig:num_comments}. 
The mean message count was 4.71 and 6.49 for functional and security bugs, respectively.
Of the 23 bugs which required more than 10 messages, 16 (69.6\%) were security bugs.
This suggests that \textbf{security bugs are more "conceptually complex" and require more discussion to identify the most appropriate corrective action compared to functional bugs}. 
The times taken to close a bug report is shown in \autoref{fig:time_to_fix}. 
The mean was 35.6 days for functional and 21.2 days for security bugs, suggesting that security bug fixes are prioritized. 
This result is consistent with the observations of prior work in software bugs~\cite{Zaman_Adams_Hassan_2011}.

\subsection{Insights into Bug Fix Characteristics}
To simplify our measurement of the bug fixes characteristics, we constrain our dataset to design files and exclude files related to documentation, scripts, DV, etc. 
OpenTitan includes several tools to auto-generate RTL, e.g., its \textit{regtool} uses .hjson files to auto-generate register design files that connect to the tlul bus. 
In bug fixes that required changes to these auto-generated files, we included the auto-generated design files but exclude the input (.hjson). 
We excluded two bug fixes with 42 and 45 files changed; both bug fixes changed the regtool itself, resulting in changes to all files it generated.

\begin{figure}[t]
    \centering
    \subfloat[Verilog continuous assignment statement]{\includegraphics[width=0.6\columnwidth]{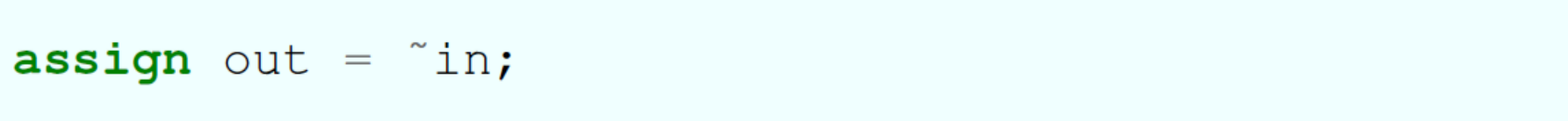}}
    \vfill
    \subfloat[AST from Slang in JSON]{\includegraphics[width=0.6\columnwidth]{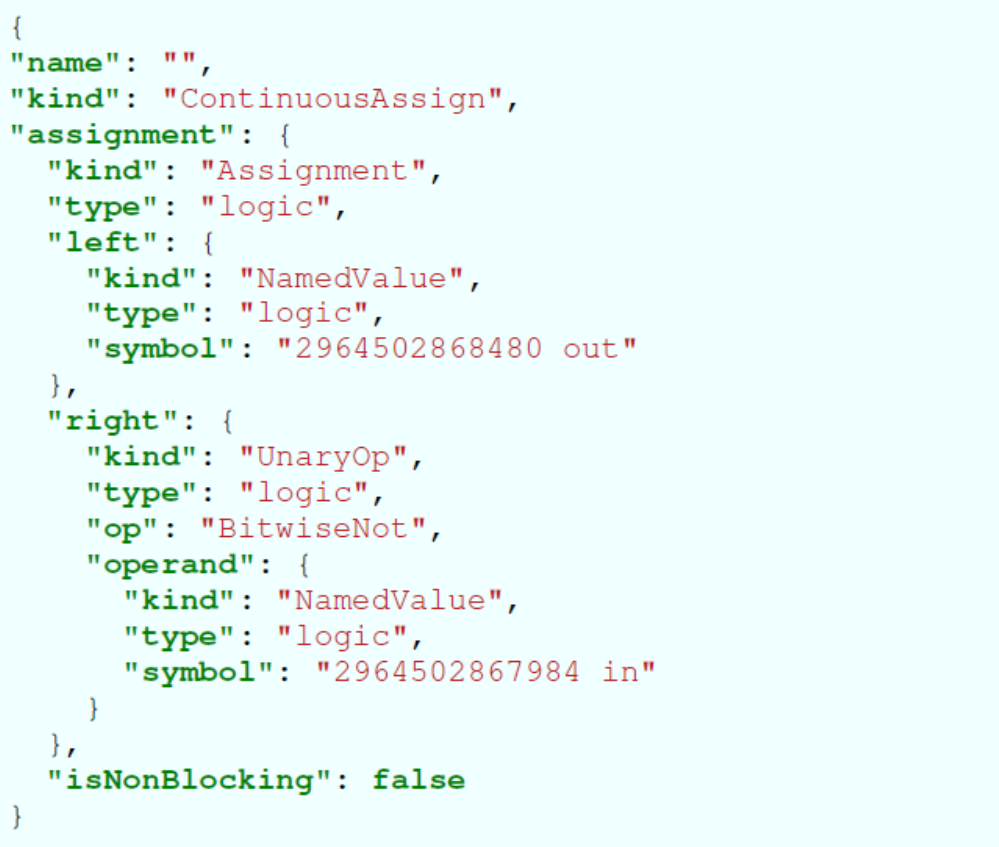}}
    \caption{Example AST from Slang\label{fig:ast}}
\end{figure}

Our analyses included an \ac{AST} comparison of design files. An \ac{AST} is a tree representation of the source code created by the parser during compilation.
It is constructed using the grammar of a formal language and expresses grammatical information through its structure (i.e., node types and connections). 
It also discards elements that are unnecessary to synthesis, such as comments. 
The tree structure facilitates the analysis of source code for semantic checking (e.g., type mismatches) and the creation of graphs for optimizations (e.g., constant propagation). 
\autoref{fig:ast} illustrates the AST created for a verilog statement using Slang\footnote{\url{https://github.com/MikePopoloski/slang}}.\\

\noindent \textbf{RQ2.a): What is the size/locality of fixes?}
\begin{figure}[t]
    \centering
    \includegraphics[width=0.95\columnwidth]{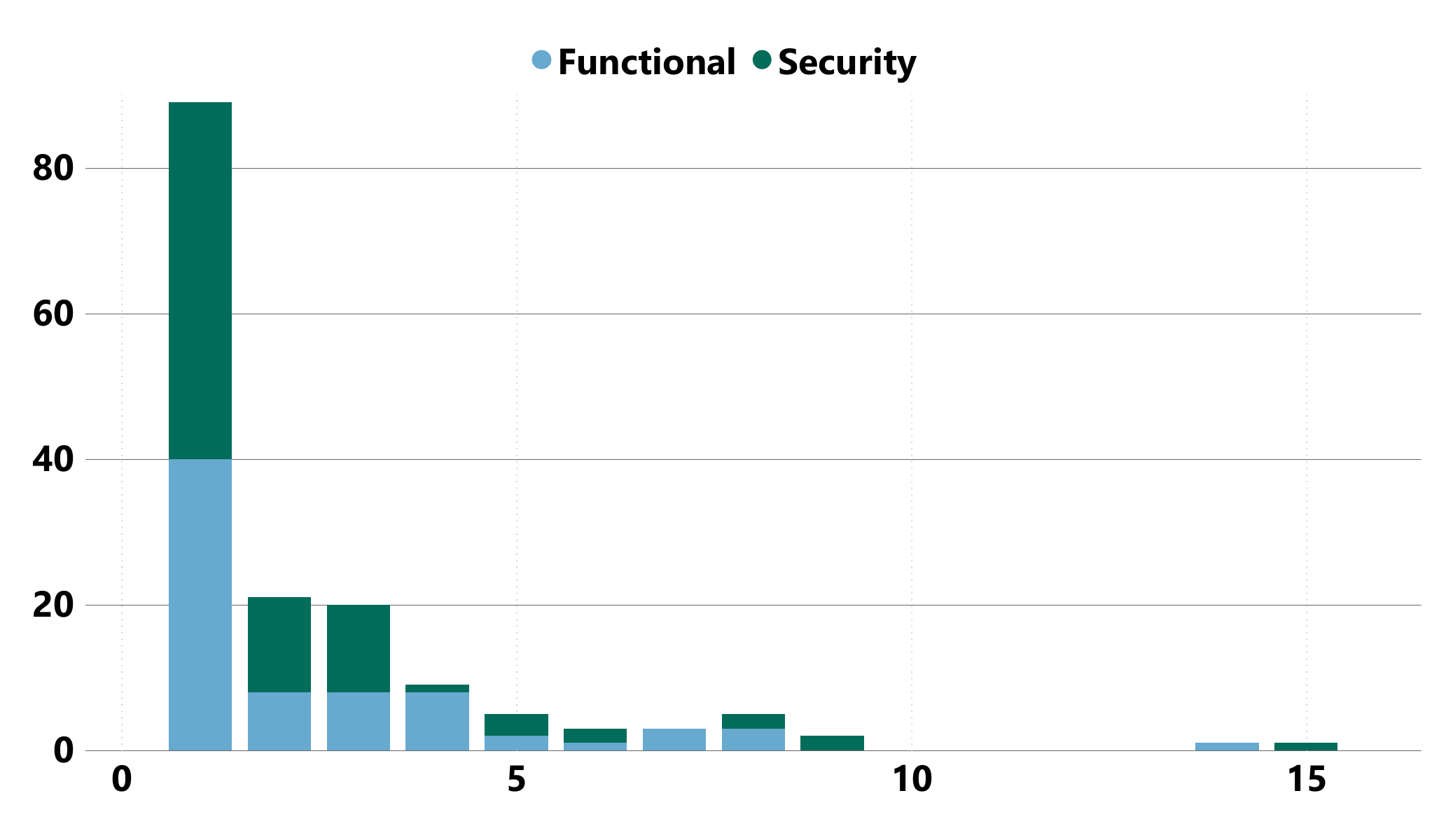}
    \vspace{-1.5em}
    \caption{\# Bugs vs. \# files changed for fixes (Distribution of the number of lines changed in bug fixes).}
    \label{fig:num_files_changed}
\end{figure}
First, we examined the number of design files changed in bug fixes, as shown in \autoref{fig:num_files_changed}. 
We found that, on average, 2.91 files are changed, with 3.13 files and 2.71 files changed for functional and security bugs, respectively. 
We also observed that \textbf{55.3\% of bug fixes only modify one file.}
Next, we inspected the total \# of lines changed for bug fixes (or ``footprint'', defined as the total number of lines added combined with the total number of lines removed). 
We found that \textbf{32.7\% and 61.0\% of bug fixes modify at most 10 lines and 30 lines, respectively}.
This insight motivates the development of detection and repair techniques that can leverage both the high locality and low footprint of hardware bugs, such as static analysis approaches proposed by Ahmad et al.~\cite{Ahmad_cweat_2022}. 

\begin{figure}[t]
    \centering
    \includegraphics[width=0.95\columnwidth]{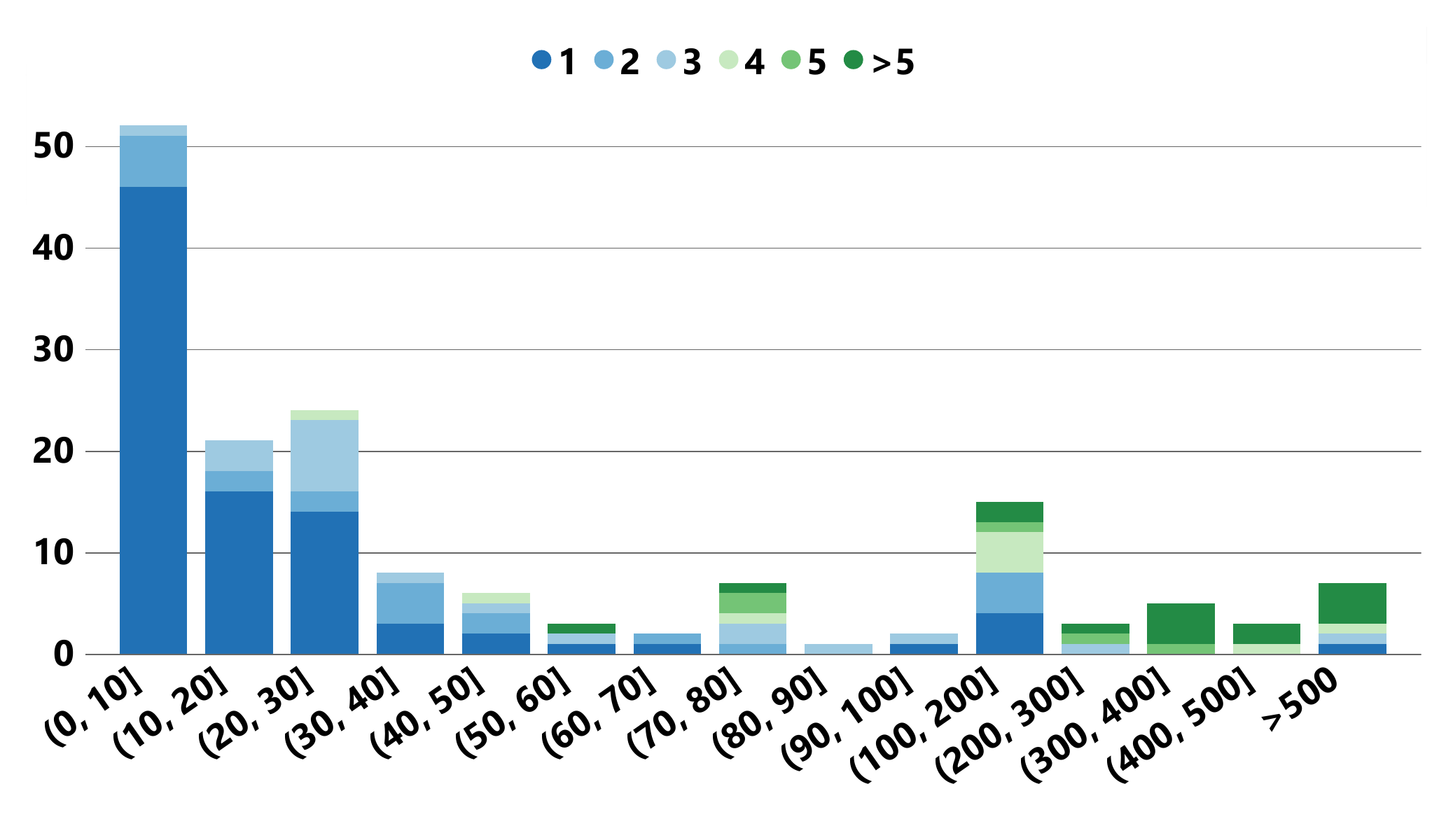}
    \vspace{-1.5em}
    \caption{\# bug fixes vs. the \# lines changed for a fix (``footprint''). Different colors represent the number of files changed in a fix (from 1 to more than 5).}
    \label{fig:lines_vs_files}
\end{figure}

Finally, we studied the relationship between the number of files and the total number of lines changed . 
~\autoref{fig:lines_vs_files} shows the number of bugs fixed by changing a given number of lines (and colored/grouped by the number of files changed). 
Bug fixes involving 10 or fewer lines changed occur in 1, 2, or 3 files. 
This suggests that bug fixes with "small footprints" (i.e., a low number of lines changed) can occur across multiple files. 
Hardware is highly hierarchical and inter-modular, and bugs that involved the integration and/or interactions between design elements likely require modifications across multiple files, even for "minor" changes.\\

\begin{figure}[t]
    \centering
    \includegraphics[width=0.95\columnwidth]{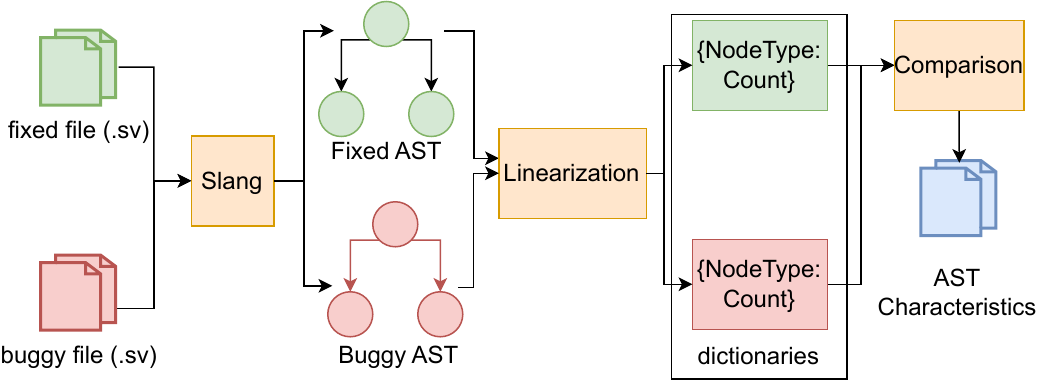}
    \caption{AST Comparison Framework}\label{fig:ast_flow}
\end{figure}

\noindent \textbf{RQ2.b): What language constructs are associated with bug fixes?}
We compared the \ac{AST} of each modified design file before and after a bug fix to gain insight into the characteristics of bug fixes from a language construct perspective.
We achieved this by linearizing the AST using node counts; we store the count of each node in the tree and compare the counts obtained from the two ASTs.
The AST comparison framework is illustrated in \autoref{fig:ast_flow}.
We started with a buggy and fixed file pair and parse both files using Slang to obtain their ASTs. 
Afterward, we traverse each tree (implementing traversal using the visitor design pattern). 
If the node being "visited" is a member (i.e., hierarchical element), statement, or assignment expression, the count of that node type is incremented in a dictionary data structure until the end of the tree is reached. 
After traversal, we compared the two ASTs by computing the difference between the dictionaries obtained from each AST. 

\begin{figure}[t]
    \centering
    \includegraphics[width=0.95\columnwidth]{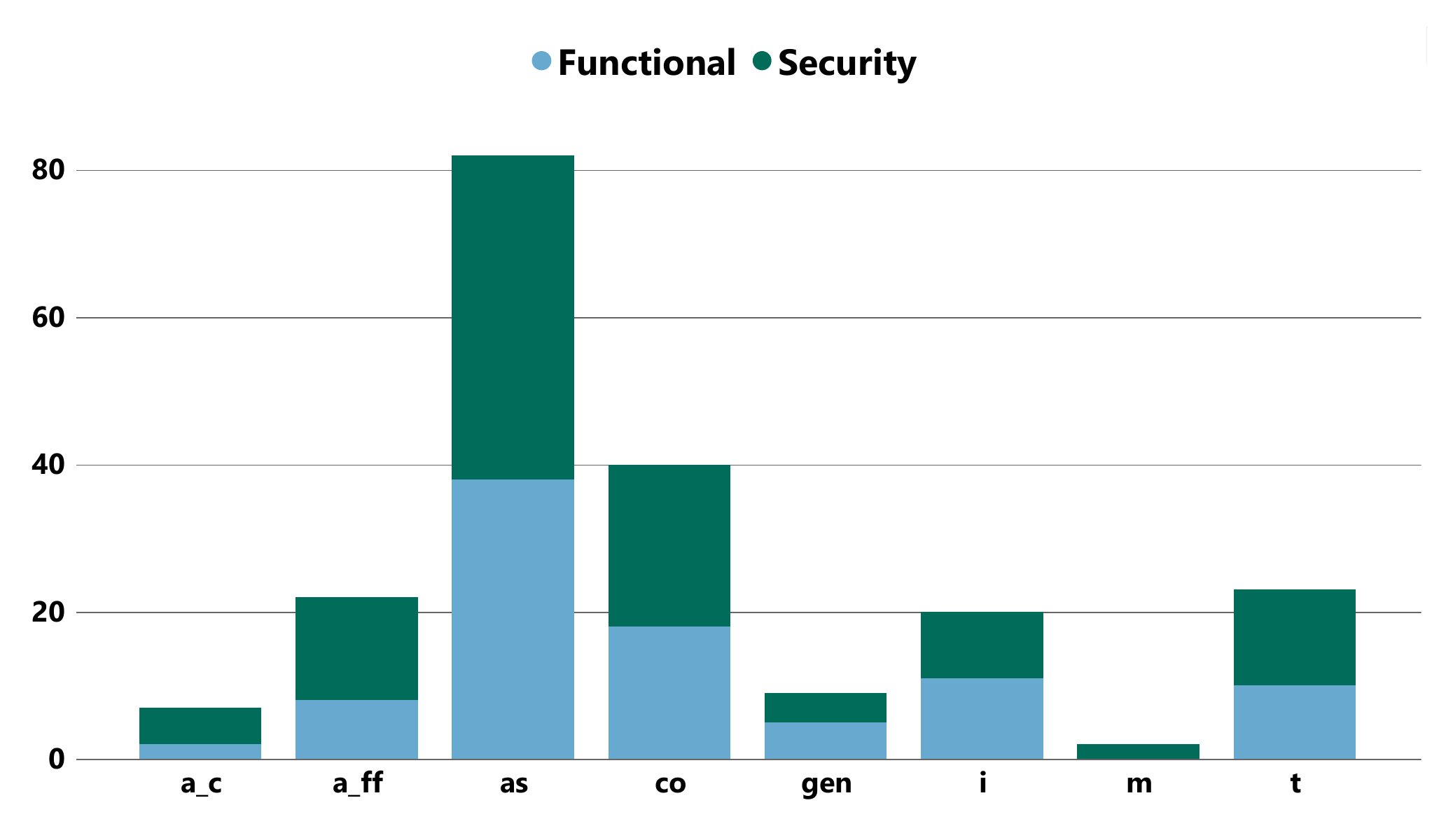}
    \vspace{-2em}
    \caption{\# bug reports containing node types of interest. Node types: assignment (as), generate (gen), always\_comb (a\_c), always\_ff (a\_ff), CaseStatement (ca), ConditionalExpression (t), ConditionalStatement (co), HierarchyInstantiation (i), ModuleDeclaration (m).}
    \vspace{-1.5em}
    \label{fig:bug_fixes_vs_node}
\end{figure}

We examined the node types that are most often involved across all bug fixes (i.e., bug fixes that added or removed at least one node of a given type). 
\autoref{fig:bug_fixes_vs_node} illustrates the number of bug fixes that contain node types of interest. 
We omit statements like data declarations as these do not provide semantic meaning. 
\textbf{Assignments, "as" (blocking and non-blocking assignments) are the most common, at 50.1\% of all bug fixes.}
The other types are as follows: conditional statements "co" (24.8\%), ternary operator "t" (14.3\%), always\_ff "a\_ff" (23.2\%), hierarchy instantiation "i" (12.4\%), module declaration "m" (1.2\%), generate statements "gen" (5.6\%), and always\_comb "a\_c" (4.3\%). 
This aligns with the ranges found by Sudhakrishnan et al. in prior work~\cite{Sudakrishnan_verilog_bugfix_patterns_2008}.
We also observe that there is no proportional difference between functional and security bug fixes across node types.
This indicates that functional and security bug fixes are similar at the structural level.

\section{Discussion and Takeaways\label{sec:discussion}} 
There are fewer sources of reliable data in open-source hardware than software.
Most open-source hardware projects are relatively new and bug reporting processes/quality is more rudimentary and do not consider security as a design goal.
This limited our analysis to the OpenTitan SoC; 
we believe that our choice of SoC is fairly representative of a comprehensive, actively developed hardware project.
Our analyses could be applied to other hardware projects, both open-source and proprietary, to gain further insights and provides a useful contemporary indication of bug characteristics. 
We open-source our analysis tools to facilitate this\footnote{\url{https://github.com/joeya20/OpenTitan-Analysis}}.

We performed manual analysis for our initial bug classification (security vs. non-security). 
This is inherently a subjective approach and it is possible that others may disagree with some of our choices. 
We attempted to minimize bias by completing our classification over multiple iterations.
During our analyses, we encountered several instances that we found were not clear-cut (we thus took a conservative posture and marked security-relevant bugs when we were highly confident), which reiterates the broader challenge of security-related reasoning for hardware designs. 

We summarize our key findings below:
\begin{itemize}
    \item 52.9\% of the bugs found in OpenTitan in the period studied were security bugs, required more discussion to fix, and were fixed faster than functional bugs.
    \item cryptography IPs contained the most security bugs and had a higher correlation with confidentiality than other impacts.
    \item Most bug fixes modify only one file and at most 30 lines with no major quantitative difference between functional and security bugs.
    \item Assignments were involved in half of all bug repairs.
\end{itemize}

These insights provide valuable takeaways and future direction in hardware security verification. 
Our results suggest that future bug repair techniques should favor high locality and low footprint repairs, with a bias towards assignment-related modifications. 
They also indicate that although cryptography IPs are likely heavily designed and scrutinized due to their security-critical purpose, they are where most security bugs are found. 
We also observed no appreciable quantitative differences between functional and security bug fixes, indicating that automated bug classification should focus on qualitative features that better capture bug context.
\section{Conclusions\label{sec:conclusions}}
To gain new insights into hardware security bugs in the HDL level, we investigated open-source hardware projects. 
Through a deep dive into OpenTitan project's bug reports and fixes involving manual analysis and an abstract syntax tree-based comparison, we identified several characteristics of hardware security bugs and fixes.
We found that 52.9\% of bugs in 2018 to 2022 were security bugs, most of which occurred in cryptography-related IPs. 
Futhermore, 55.3\% of bug fixes only modify one file, with assignment modifications being most common across fixes. 
Our results provide new insights that can inform future work in HDL static security analysis and repair tools. 
Our future work will expand our analyses to more projects and investigate the design of static analysis scanners that can better handle potential bugs across different files, focusing on prioritizing bug searching based on AST node types. 


\section*{Acknowledgments}
We would like to acknowledge CMC Microsystems for the provision of products and services that facilitated this research. 
This research work is supported in part by a gift from Intel Corporation. This work does not in any way constitute an Intel endorsement of a product or supplier.
We acknowledge the support of the Natural Sciences and Engineering Research Council of Canada (NSERC), RGPIN-2022-03027. 
Joey was supported by an NSERC Undergraduate Student Research Award (USRA) for part of this work. 

\bibliographystyle{IEEEtran}
\bibliography{IEEEabrv,references}

\end{document}